\documentclass[11pt]{article}

\usepackage[margin=1in]{geometry}
\usepackage{graphicx}
\usepackage{booktabs}
\usepackage{array}
\usepackage{amsmath}
\usepackage{amssymb}
\usepackage{microtype}
\usepackage{natbib}
\usepackage{hyperref}
\usepackage{xcolor}
\usepackage{url}
\usepackage{xurl}
\usepackage{caption}
\usepackage{subcaption}
\usepackage{enumitem}
\usepackage{placeins}

\hypersetup{
  colorlinks=true,
  linkcolor=blue,
  citecolor=blue,
  urlcolor=blue
}

\newcommand{\pkg}[1]{\texttt{#1}}
\newcommand{\code}[1]{\nolinkurl{#1}}

\title{
Contrast-Aware Annotation for Statistical Graphics:\\
The \pkg{ggtwotone} Package for \pkg{ggplot2}
}

\author{
Beenu Sareena$^{*}$ \qquad Heike Hofmann\\[6pt]
\small Department of Statistics, University of Nebraska--Lincoln\\
\small Nebraska, USA\\[3pt]
\small \texttt{79365709@nebraska.edu} \qquad
\texttt{hhofmann4@unl.edu}\\[3pt]
\small $^{*}$Corresponding author
}

\date{}

\begin{document}

\maketitle

% ============================================================
% Abstract
% ============================================================

\begin{abstract}

Readable annotations are essential for interpreting statistical
graphics, yet conventional single-color lines and labels can lose
visibility when they cross backgrounds with heterogeneous luminance.
We introduce a contrast-aware annotation framework implemented in the
\pkg{ggtwotone} R package for \pkg{ggplot2}. The framework combines
dual-stroke rendering, adaptive text-color selection, and perceptually
guided highlight palettes to improve annotation visibility across light
and dark regions. A shared contrast-adjustment engine supports WCAG- and
APCA-based contrast criteria, reducing the need for manual color
adjustment. The package provides contrast-aware geoms for segments,
curves, paths, mathematical functions, regression overlays, and text
while remaining compatible with standard \pkg{ggplot2} workflows.

A simulation-based evaluation across heterogeneous background colors
demonstrates improved worst-case contrast for dual-stroke annotations
and adaptive text selection, while also identifying trade-offs in
perceptual separation as the number of requested highlight colors
increases. Applications to statistical graphics and scientific images
illustrate the framework in practical visualization settings.
Together, these tools provide a reproducible approach for incorporating
contrast considerations directly into graphical annotation.

\end{abstract}

\noindent\textbf{Keywords:}
statistical graphics; contrast-aware annotation; dual-stroke rendering;
perceptual contrast; visualization accessibility; ggplot2; R

% ============================================================
\section{Introduction}
% ============================================================

Statistical graphics are fundamental tools for exploring, communicating,
and interpreting data, often relying on annotations such as reference
lines, regression curves, arrows, labels, and highlighted regions to
guide interpretation. These graphical elements help readers identify
important patterns, compare groups, and connect visual features with
statistical conclusions. Their effectiveness, however, depends on
remaining visually distinguishable from the surrounding figure. When
annotations become difficult to perceive, the interpretability of the
visualization is reduced, regardless of the quality of the underlying
analysis. Annotation readability is therefore not merely an aesthetic
consideration but an important component of effective statistical
communication.

Maintaining annotation visibility is particularly challenging when
graphical elements traverse heterogeneous backgrounds. Modern
statistical graphics frequently combine multiple colors, gradients,
raster images, heatmaps, or dense collections of plotted observations
within a single figure. In these settings, a foreground color that
provides excellent contrast in one region may become nearly
indistinguishable in another. Dark annotations may disappear against
dark backgrounds, while light annotations may become difficult to
perceive over brighter regions.

Similar challenges arise in scientific image annotation, geographic
maps, microscopy, medical imaging, technical diagrams, and other
visualizations where local background characteristics vary
substantially. Consequently, producing consistently readable figures
often requires repeated manual adjustments that are time-consuming,
subjective, and difficult to reproduce.

A variety of strategies can be used to improve the visibility of
graphical annotations. Common approaches include manually selecting
foreground colors, increasing line width, adding transparency,
surrounding text with bounding boxes or halos, and adjusting individual
graphical elements after inspecting the final figure. Although these
techniques can improve readability in specific situations, they
generally require iterative manual refinement and often depend on the
particular background over which an annotation is displayed. As figures
become more complex or are reused across different media, these manually
chosen settings may no longer provide sufficient visual separation.

The increasing emphasis on accessibility further highlights the need
for principled approaches to annotation design. Scientific figures are
viewed on displays with different luminance characteristics, projected
during presentations, embedded within web pages, and reproduced in
print. Accessibility guidelines such as the Web Content Accessibility
Guidelines (WCAG) provide quantitative recommendations for evaluating
color contrast \citep{WCAG}, while newer perceptual approaches,
including the Advanced Perceptual Contrast Algorithm (APCA), have
motivated reconsideration of how contrast criteria developed for text
and interface elements may relate to visualization \citep{Muth2025}.

In this paper, we introduce a contrast-aware annotation framework based
on dual-tone rendering. Rather than relying on a single foreground
color, line-based graphical elements are represented using
complementary light and dark strokes. This representation reduces the
dependence of annotation visibility on a single local background color.
The same contrast-aware principle is extended to adaptive text rendering
and context-dependent highlight palette generation.

The Grammar of Graphics implemented in \pkg{ggplot2} provides a layered
and extensible framework for statistical visualization
\citep{ggplot2}. Extending this framework provides an opportunity to
incorporate perceptual principles directly into the visualization
pipeline while preserving familiar syntax. Rather than introducing a
separate graphics system, our approach integrates with existing
\pkg{ggplot2} workflows.

We present \pkg{ggtwotone}, an R package that implements this
contrast-aware annotation framework through a collection of geoms and
supporting utilities. The package provides dual-stroke geoms for line
segments, curves, paths, mathematical functions, and regression
overlays, together with adaptive text rendering and perceptually guided
highlight palette generation. A common contrast-adjustment engine
supports these components using WCAG- and APCA-based contrast
evaluation.

The main contributions of this work are:

\begin{enumerate}[label=(\arabic*)]

\item a general contrast-aware framework for annotations over
heterogeneous backgrounds;

\item dual-stroke rendering methods for segments, curves, paths,
mathematical functions, and regression overlays;

\item adaptive text rendering that selects foreground colors according
to local background contrast;

\item context-dependent highlight palette generation combining contrast
and perceptual color-separation criteria;

\item an implementation of these methods in the \pkg{ggtwotone} package
that integrates with the \pkg{ggplot2} layer system; and

\item a simulation-based evaluation characterizing the contrast
behavior and practical trade-offs of the proposed methods.

\end{enumerate}

The remainder of this paper reviews related work and introduces the
design principles underlying \pkg{ggtwotone}. We then describe the
contrast-aware rendering framework and its implementation, demonstrate
the package through representative applications, and evaluate its
behavior through simulation. We conclude with a discussion of the
framework's strengths, limitations, and opportunities for future
development.

% ============================================================
\section{Related Work}
% ============================================================

Color selection and perceptual accessibility have long been important
considerations in statistical graphics. Zeileis, Hornik, and Murrell
\citep{Zeileis2009} advocated the use of perceptually based HCL color
spaces for statistical graphics, emphasizing the importance of
controlling hue, chroma, and luminance when constructing qualitative and
sequential color schemes.

Related work has considered the accessibility of scientific color maps
for viewers with color-vision deficiencies. For example, Geissbuehler
and Lasser \citep{Geissbuehler2013} developed color schemes designed to
remain interpretable under red--green color-perception deficiencies,
while Nuñez, Anderton, and Renslow \citep{Nunez2018} proposed the
perceptually uniform \emph{cividis} colormap with similar accessibility
objectives. These approaches primarily address colors used to encode
data rather than the visibility of annotations placed over
heterogeneous backgrounds.

Accessibility has also become an explicit consideration in software for
statistical graphics. The \pkg{ggpubfigs} package provides
colorblind-friendly palettes and themes for producing accessible,
publication-quality \pkg{ggplot2} figures \citep{Steenwyk2021}, while
tools such as Color Quest support the evaluation and exploration of
palettes under different forms of color-vision deficiency
\citep{Nelli2024}.

More generally, accessibility guidance recommends avoiding reliance on
color alone and using differences in lightness or redundant visual
encodings when appropriate. WCAG provides quantitative contrast criteria
for distinguishing foreground content from its background
\citep{WCAG}, and more recent visualization discussions have considered
how such criteria and newer perceptual approaches, including APCA,
relate to data visualization \citep{Muth2025}.

A related problem arises when text or other foreground elements are
placed over visually complex backgrounds. Research on typography over
complex imagery has shown that legibility depends strongly on the
interaction between foreground content and the underlying background,
and has examined techniques such as scrims, blur, and drop shadows for
improving separation \citep{Sawyer2020}. Such methods illustrate a
broader principle relevant to statistical graphics: a foreground
treatment that works well in one region may not remain effective when
local background luminance or texture changes.

Within R, \pkg{ggplot2} provides an extensible implementation of the
Grammar of Graphics in which graphical elements are constructed as
layers \citep{ggplot2}. This architecture has enabled a large ecosystem
of extensions for specialized graphical tasks. Existing accessibility
approaches have largely focused on selecting suitable palettes,
simulating color-vision deficiencies, or modifying overall figure
appearance.

These approaches are complementary to, but distinct from, the problem
addressed here: maintaining the local visibility of annotations whose
geometry crosses backgrounds with changing luminance or color.

\pkg{ggtwotone} addresses this problem by incorporating contrast
adaptation directly into annotation rendering. Rather than requiring a
single globally suitable foreground color, the package combines
complementary dual-stroke rendering for line-based annotations,
adaptive foreground selection for text, and context-dependent highlight
palette generation within a common contrast framework. The contribution
is therefore not an alternative to perceptually designed or
colorblind-friendly palettes. Instead, it complements these approaches
by addressing the local foreground--background relationship encountered
by annotations in statistical graphics and scientific images.

% ============================================================
\section{Design Principles}
% ============================================================

The development of \pkg{ggtwotone} was guided by a small set of design
principles intended to improve annotation readability while preserving
the flexibility and reproducibility of the \pkg{ggplot2} ecosystem.
Rather than treating annotation visibility as a problem to be addressed
manually after a figure has been created, the package incorporates
contrast considerations directly into the rendering process.

% ------------------------------------------------------------
\subsection{Preserve Annotation Visibility}
% ------------------------------------------------------------

The primary objective of the package is to improve the visibility of
graphical annotations across heterogeneous backgrounds. Conventional
annotations rely on a single foreground color, making their readability
dependent on local background characteristics. As an annotation
traverses regions of varying luminance, a color that is clearly
distinguishable in one region may become difficult to perceive in
another.

Rather than attempting to identify a universally optimal foreground
color, \pkg{ggtwotone} represents line-based annotations using
complementary light and dark strokes. This design increases the
likelihood that at least one component maintains useful contrast as the
local background changes.

% ============================================================
% FIGURE 1
% ============================================================

\begin{figure}[htbp]
\centering

\includegraphics[
   width=0.85\textwidth,
  keepaspectratio
 ]{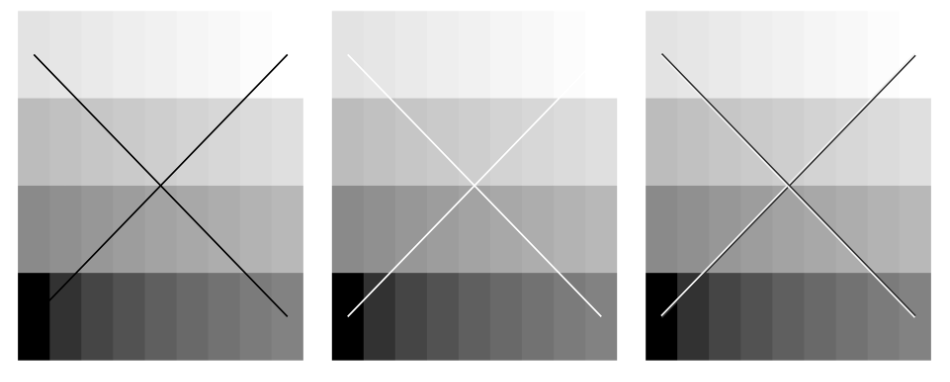}

\caption{Comparison of annotation visibility across a heterogeneous
grayscale background using a single black stroke, a single white
stroke, and dual-stroke rendering.}

\label{fig:segment-comparison}
\end{figure}

% ------------------------------------------------------------
\subsection{Integrate Accessibility into Rendering}
% ------------------------------------------------------------

Accessibility considerations are most reproducible when incorporated
during figure construction rather than evaluated only after a
visualization has been completed. The package therefore supports
contrast evaluation using WCAG and APCA criteria. By incorporating
contrast evaluation into annotation construction, the resulting
foreground choices become reproducible properties of the visualization
rather than solely manual design decisions.

% ------------------------------------------------------------
\subsection{Maintain Compatibility with the Grammar of Graphics}
% ------------------------------------------------------------

A central design goal was to integrate with existing \pkg{ggplot2}
workflows and its layered Grammar of Graphics framework. Rather than
introducing a new graphics system or requiring users to adopt unfamiliar
syntax, the package provides contrast-aware alternatives to familiar
layers while preserving the general aesthetics and layer-based
philosophy of \pkg{ggplot2}.

% ------------------------------------------------------------
\subsection{Minimize Manual Intervention}
% ------------------------------------------------------------

Producing publication-quality figures can require repeated adjustments
to annotation colors, line widths, and label styles after inspecting
the completed visualization. Such iterative tuning is time-consuming
and can be difficult to reproduce.

Several package functions automate portions of these decisions from
user-specified inputs. These include \code{adjust_contrast_pair()},
\code{geom_text_contrast()}, and \code{highlight_colors()}.

% ------------------------------------------------------------
\subsection{Provide a Unified Contrast-Aware Framework}
% ------------------------------------------------------------

Although individual package components address different visualization
tasks, they share a common objective and computational foundation.
Dual-stroke rendering, adaptive text-color selection, and highlight
palette generation use related contrast-evaluation principles. This
common framework promotes consistency across annotation types and
provides a basis for future extensions.

% ============================================================
\section{Contrast-Aware Rendering Framework}
\label{sec:framework}
% ============================================================

The contrast-aware rendering framework underlying \pkg{ggtwotone} is
organized around a common objective: graphical annotations should remain
distinguishable as the local background changes. The framework combines
quantitative contrast evaluation with complementary foreground colors,
dual-stroke rendering, adaptive text selection, and perceptually
constrained highlight generation.

Let $B(u)$ denote the background color encountered by an annotation at
position $u$, and let $\mathcal{C}(c,B(u))$ denote the contrast between
foreground color $c$ and the local background. In \pkg{ggtwotone},
$\mathcal{C}$ may be evaluated using either WCAG or APCA criteria. This
notation provides a common basis for describing the dual-stroke, text,
and highlight-selection methods below.

% ------------------------------------------------------------
\subsection{Contrast Measurement}
% ------------------------------------------------------------

Contrast provides the quantitative foundation for color selection in
\pkg{ggtwotone}. For a foreground color $c$ displayed over background
$B$, let

\[
\mathcal{C}(c,B)
\]

denote their contrast according to the selected criterion.

Under the WCAG formulation, contrast is computed from relative
luminance. For two colors with relative luminances $L_1$ and $L_2$,
where $L_1 \geq L_2$, the contrast ratio is

\begin{equation}
\operatorname{CR}(c_1,c_2)
=
\frac{L_1+0.05}{L_2+0.05}.
\label{eq:wcag}
\end{equation}

The resulting ratio ranges from 1 for equal relative luminance to 21
for the maximum black--white contrast. APCA provides an alternative,
polarity-sensitive measure of lightness contrast. Because WCAG and APCA
operate on different numerical scales, \pkg{ggtwotone} treats the
selected contrast method and its corresponding threshold jointly rather
than interpreting their raw values as interchangeable.

The function \code{adjust\_contrast\_pair()} uses the selected contrast
criterion to derive complementary light and dark variants from a
user-specified base color. These colors provide the foundation for the
dual-stroke and adaptive annotation methods described below.

% ------------------------------------------------------------
\subsection{Dual-Stroke Representation}
% ------------------------------------------------------------

The central idea of dual-stroke rendering is to replace a single
foreground color with a complementary pair consisting of a light color
$c_L$ and a dark color $c_D$. These colors are derived from a
user-specified base color through \code{adjust\_contrast\_pair()}.

Consider an annotation parameterized by $u\in[0,1]$, with local
background color $B(u)$. At each position, define the effective
visibility margin of the dual-stroke representation as

\begin{equation}
M(u)
=
\max
\left\{
\mathcal{C}\!\left(c_L,B(u)\right),
\mathcal{C}\!\left(c_D,B(u)\right)
\right\}.
\label{eq:visibility}
\end{equation}

The use of the maximum reflects the central design principle of the
framework: local visibility is supported when at least one component of
the paired representation provides sufficient contrast with the
background.

A conservative summary over the entire annotation is

\begin{equation}
M_{\min}
=
\min_{u\in[0,1]} M(u).
\label{eq:minvisibility}
\end{equation}

Given a target contrast threshold $\tau$, the desired condition is

\begin{equation}
M_{\min}\geq\tau.
\label{eq:threshold}
\end{equation}

This formulation differs from requiring both components to satisfy the
threshold simultaneously. Instead, the paired representation provides
complementary contrast: a lighter component can provide separation over
darker regions, while a darker component can provide separation over
lighter regions.

Conceptually, for line- and curve-based annotations, let

\[
\mathbf{r}(t)
=
\begin{bmatrix}
x(t)\\
y(t)
\end{bmatrix},
\qquad t\in[0,1],
\]

represent the centerline of an annotation and let $\mathbf{n}(t)$ denote
its local unit normal vector. A dual representation can be expressed
conceptually as

\begin{equation}
\mathbf{r}_{\pm}(t)
=
\mathbf{r}(t)
\pm
\delta\,\mathbf{n}(t),
\label{eq:offset}
\end{equation}

where $\delta>0$ represents a small visual separation between the two
components. Equation~\ref{eq:offset} provides a geometric
interpretation of the paired-stroke design rather than a general
statement about all possible rendering implementations.

\begin{table}[htbp]
\centering

\caption{Symbols used in the contrast-aware dual-stroke framework.}
\label{tab:dual-symbols}

\begin{tabular}{p{0.23\textwidth}p{0.67\textwidth}}

\toprule
Symbol & Description\\
\midrule

$B(u)$ &
Local background color encountered by the annotation at position $u$\\

$c_0$ &
User-specified base annotation color\\

$c_L$ &
Light contrast-adjusted color derived from $c_0$\\

$c_D$ &
Dark contrast-adjusted color derived from $c_0$\\

$\mathcal{C}(c,B)$ &
Contrast between foreground $c$ and background $B$\\

$M(u)$ &
Local visibility margin of the dual-stroke representation\\

$M_{\min}$ &
Minimum visibility margin across the annotation\\

$\tau$ &
Target contrast threshold\\

$\mathbf{r}(t)$ &
Centerline of a segment, curve, path, function, or fitted line\\

$\mathbf{n}(t)$ &
Unit normal vector associated with $\mathbf{r}(t)$\\

$\delta$ &
Conceptual separation between paired strokes\\

$\mathbf{r}_{+}(t),\mathbf{r}_{-}(t)$ &
Conceptual paired trajectories forming the dual representation\\

\bottomrule

\end{tabular}
\end{table}

% ------------------------------------------------------------
\subsection{Contrast-Aware Text}
% ------------------------------------------------------------

Text annotations require a different treatment because duplicating
glyph geometry may reduce legibility. Instead, \pkg{ggtwotone} uses
adaptive foreground selection.

For a label with base color $c_0$ and local background $B$, the contrast
engine generates light and dark candidate colors, denoted $c_L$ and
$c_D$, and selects the candidate providing the greater contrast with
the background:

\begin{equation}
c^{*}(B)
=
\underset{c\in\{c_L,c_D\}}{\arg\max}
\;
\mathcal{C}(c,B).
\label{eq:textselection}
\end{equation}

When a target contrast threshold $\tau$ is specified, the desired
condition for the selected foreground is

\begin{equation}
\mathcal{C}\!\left(c^{*}(B),B\right)\geq\tau.
\label{eq:textthreshold}
\end{equation}

The selection can be performed independently for labels associated with
different background colors, allowing foreground color to adapt to
local conditions within the same figure. The implementation in
\code{geom\_text\_contrast()} preserves the general role of
\code{geom\_text()} while adding contrast-aware foreground selection.

% ------------------------------------------------------------
\subsection{Highlight Color Selection}
% ------------------------------------------------------------

Highlight color generation requires balancing several perceptual
constraints. A useful highlight should remain distinguishable from its
background and base graphical color and, when multiple highlights are
requested, should remain sufficiently different from the other selected
colors.

Let $\mathcal{H}$ denote a candidate set generated in HCL space. For
candidate color $h$, background color $b$, and base color $c_0$, define

\begin{equation}
\mathcal{F}
=
\left\{
h\in\mathcal{H}:
\begin{array}{l}
\mathcal{C}(h,b)\geq\tau_{\mathrm{bg}},\\[2pt]
\mathcal{C}(h,c_0)\geq\tau_{\mathrm{base}},\\[2pt]
\Delta E_{00}(h,c_0)\geq\delta_E
\end{array}
\right\},
\label{eq:highlight-feasible}
\end{equation}

where $\tau_{\mathrm{bg}}$ and $\tau_{\mathrm{base}}$ denote requested
contrast thresholds and $\Delta E_{00}$ denotes the CIEDE2000
perceptual color-difference measure \citep{Luo2001}.

When multiple highlight colors are requested, the algorithm additionally
seeks pairwise separation:

\begin{equation}
\Delta E_{00}(h_i,h_j)\geq\delta_E,
\qquad i\neq j,
\label{eq:deltaE}
\end{equation}

and a minimum circular hue separation

\begin{equation}
d_H(H_i,H_j)\geq h_{\min},
\label{eq:hue}
\end{equation}

where $H_i$ and $H_j$ denote hue angles and $d_H$ denotes circular hue
distance.

When the requested combination of constraints cannot be satisfied,
\code{highlight\_colors()} can use a controlled relaxation procedure.
Consequently, Equations~\ref{eq:highlight-feasible}--\ref{eq:hue}
describe the target constraints rather than a guarantee that every
returned palette satisfies the original thresholds without relaxation.

\begin{table}[htbp]
\centering

\caption{Symbols used in highlight color selection.}
\label{tab:highlight-symbols}

\begin{tabular}{p{0.25\textwidth}p{0.65\textwidth}}

\toprule
Symbol & Description\\
\midrule

$\mathcal{H}$ &
Candidate highlight colors generated in HCL space\\

$\mathcal{F}$ &
Candidate set satisfying the specified perceptual constraints\\

$h$ &
Candidate highlight color\\

$b$ &
Fixed background color\\

$c_0$ &
Base graphical or text color\\

$\tau_{\mathrm{bg}}$ &
Requested contrast threshold relative to the background\\

$\tau_{\mathrm{base}}$ &
Requested contrast threshold relative to the base color\\

$\Delta E_{00}$ &
CIEDE2000 perceptual color difference\\

$\delta_E$ &
Requested minimum perceptual color separation\\

$H_i,H_j$ &
Hue angles of selected highlight colors\\

$d_H$ &
Circular distance between hue angles\\

$h_{\min}$ &
Requested minimum pairwise hue separation\\

\bottomrule

\end{tabular}
\end{table}

Together, these formulations provide a common mathematical basis for
the three principal components of \pkg{ggtwotone}. Dual-stroke geoms
use complementary colors and graphical redundancy, contrast-aware text
selects foreground colors according to local background conditions, and
highlight palette generation extends the same contrast principle with
additional perceptual-separation constraints.

% ============================================================
\section{Software Design and Implementation}
% ============================================================

The \pkg{ggtwotone} package extends the Grammar of Graphics implemented
by \pkg{ggplot2} by introducing contrast-aware annotation layers while
preserving the familiar syntax and workflow of existing graphics.
Rather than providing a separate visualization system, the package
implements new geoms and supporting utilities that integrate with the
standard \pkg{ggplot2} rendering pipeline.

% ------------------------------------------------------------
\subsection{Package Architecture}
% ------------------------------------------------------------

The package is organized around a shared contrast engine that supports
multiple annotation components. The first component consists of
dual-stroke geoms that provide contrast-aware alternatives for line
segments, curves, paths, mathematical functions, and regression
overlays.

The second component contains contrast-aware annotation utilities,
including automatic text-color selection through
\code{geom\_text\_contrast()} and perceptually guided highlight
generation through \code{highlight\_colors()}.

The third component provides shared contrast evaluation and color
adjustment using WCAG- or APCA-based criteria. Centralizing this
functionality promotes consistent behavior across package components.

% ============================================================
% FIGURE 2: Package architecture
% ============================================================

\begin{figure}[htbp]
\centering

\includegraphics[
  width=0.90\textwidth,
  keepaspectratio
]{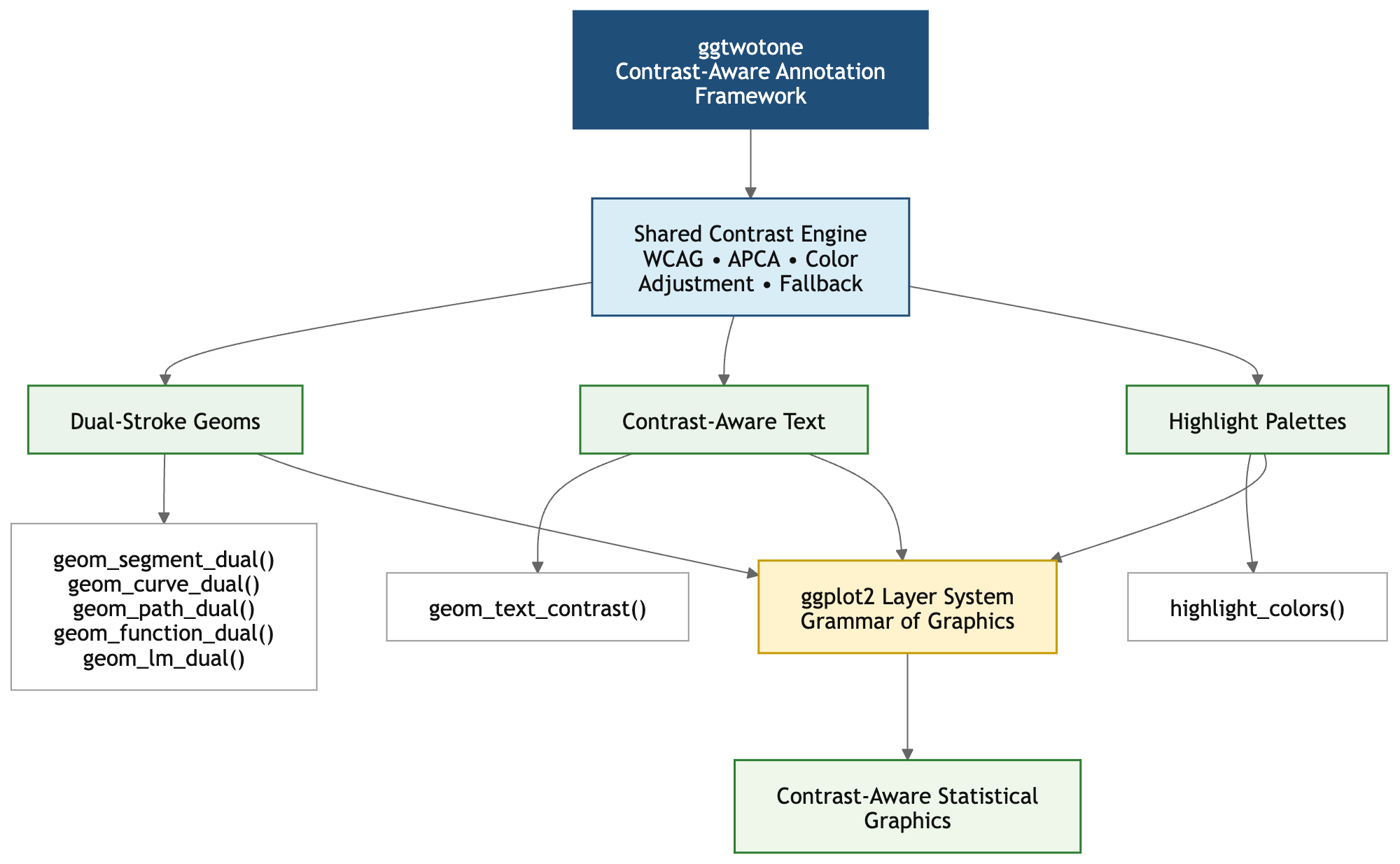}

\caption{Architecture of the \pkg{ggtwotone} package. A shared contrast
engine provides common functionality for dual-stroke geoms,
contrast-aware text rendering, and highlight palette generation. These
components integrate with the \pkg{ggplot2} layer system.}

\label{fig:architecture}
\end{figure}

% ------------------------------------------------------------
\subsection{Shared Contrast Engine}
% ------------------------------------------------------------

Contrast-aware operations within the package are supported by a shared
contrast-adjustment engine centered on
\code{adjust\_contrast\_pair()}. Rather than allowing each geom to
implement an independent color-selection strategy, this function
provides a common mechanism for deriving light and dark candidates from
a user-specified base color under the selected contrast criterion.

\begin{figure}[p]
\centering

\includegraphics[
  width=0.92\textwidth,
  height=0.88\textheight,
  keepaspectratio
]{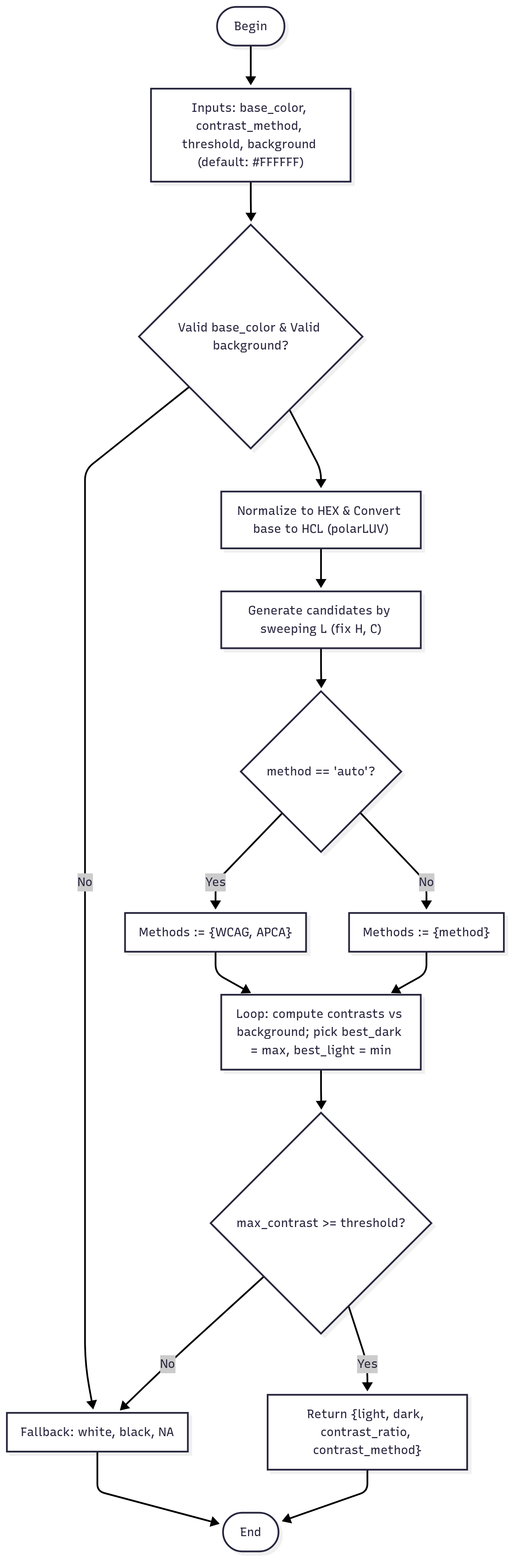}

\caption{Workflow of the shared contrast-adjustment procedure used to
derive contrast-aware light and dark color candidates from a
user-specified base color.}

\label{fig:contrast-engine-flow}
\end{figure}

Centralizing contrast functionality has several advantages. It promotes
consistent color behavior across package components, allows
improvements to the contrast-selection procedure to benefit multiple
geoms, and provides a common foundation for future extensions. The
workflow of the shared contrast-adjustment engine is summarized in
Figure~\ref{fig:contrast-engine-flow}.

% ============================================================
% FIGURE 3: Contrast engine workflow
% ============================================================

\clearpage

\clearpage

% ------------------------------------------------------------
\subsection{Dual-Geometry Rendering}
% ------------------------------------------------------------

The dual-stroke geoms share a rendering strategy in which a graphical
primitive is represented using complementary light and dark components.
The paired representation provides graphical redundancy across
heterogeneous backgrounds: when one component has low contrast with a
local region, the other may retain greater contrast.

This strategy differs from relying only on increased line width or
transparency because visibility is supported through complementary
foreground components rather than modification of a single foreground
color.

% ------------------------------------------------------------
\subsection{Integration with \pkg{ggplot2}}
% ------------------------------------------------------------

Compatibility with existing \pkg{ggplot2} workflows was a primary
design objective. Whenever possible, functions were designed as
contrast-aware alternatives to familiar layers while preserving
aesthetic mappings and layer semantics.

For example, \code{geom_lm_dual()} provides a contrast-aware regression
overlay. The function \code{geom_text_contrast()} extends the role of
\code{geom_text()} by adding foreground adjustment. Users can therefore
introduce contrast-aware annotations incrementally without restructuring
the remainder of an existing plot.

% ============================================================
\section{Applications}
% ============================================================

The previous sections described the perceptual principles, mathematical
framework, and software architecture underlying \pkg{ggtwotone}. We now
illustrate how these components operate through representative
visualization examples. The objective is not to alter the statistical
information represented by the figures, but to improve the visibility
of annotations across varying local background conditions.

% ------------------------------------------------------------
\subsection{Dual-Stroke Segments and Curves}
% ------------------------------------------------------------

Dual-stroke segments and curves are useful when annotations cross
regions with substantially different luminance. In a conventional
single-color annotation, the selected foreground may be clearly visible
over one portion of the background but difficult to distinguish
elsewhere.

\begin{figure}[htbp]
\centering

\begin{subfigure}[t]{0.92\textwidth}
\centering

\includegraphics[
  width=\textwidth,
  keepaspectratio
]{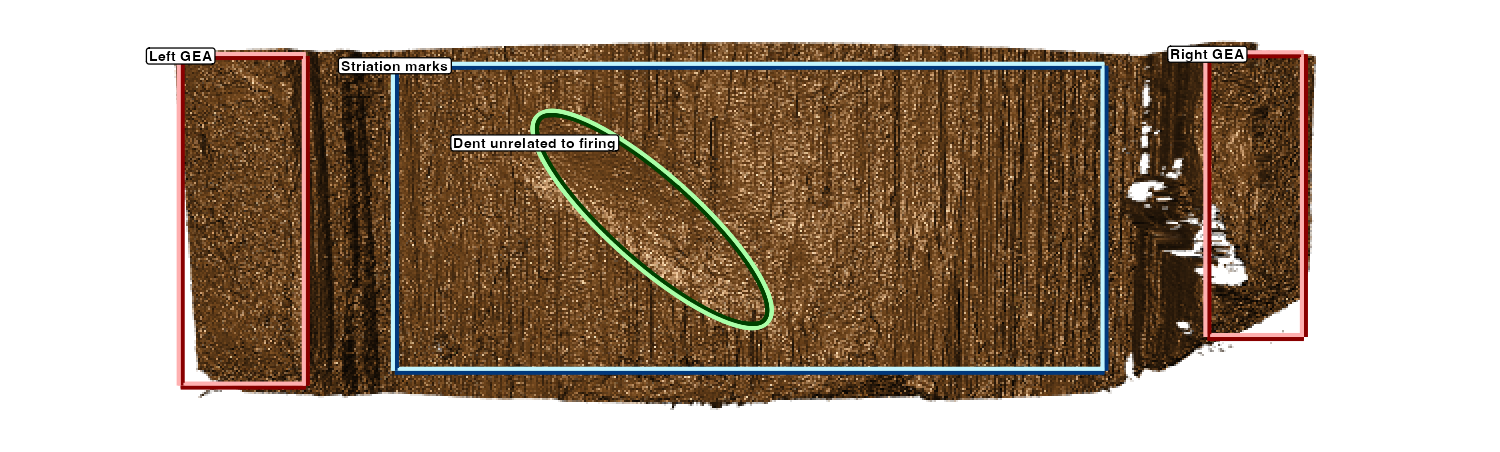}

\caption{Original rendering.}

\end{subfigure}

\vspace{0.7em}

\begin{subfigure}[t]{0.92\textwidth}
\centering

\includegraphics[
  width=\textwidth,
  keepaspectratio
]{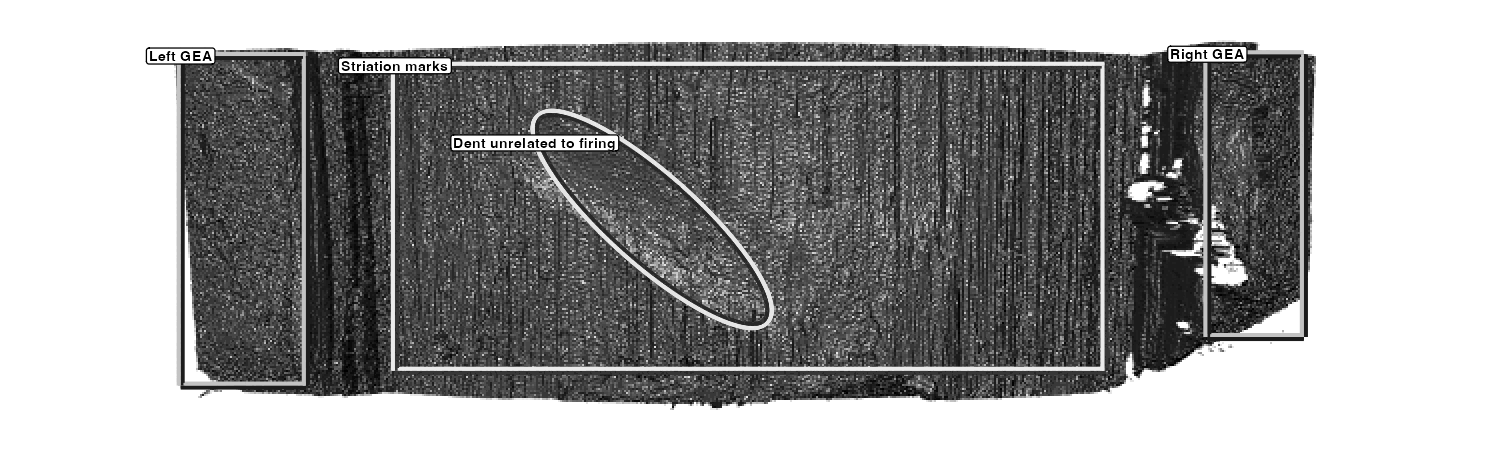}

\caption{Grayscale rendering.}

\end{subfigure}

\caption{Dual-stroke segment and curve annotations applied to a 3D
bullet land surface image, shown in the original rendering and after
grayscale conversion. Bullet surface image adapted from
\citet{hofmann_bullet_image}.}

\label{fig:bullet}

\end{figure}

The functions \code{geom\_segment\_dual()} and
\code{geom\_curve\_dual()} address this problem using complementary
light and dark components. The scientific-image example in
Figure~\ref{fig:bullet} demonstrates this behavior on a bullet surface
containing substantial spatial variation in texture and brightness.

% ============================================================
% FIGURE 4: Bullet surface
% ============================================================

\begin{figure}[htbp]
\centering

\includegraphics[
  width=0.70\textwidth,
  keepaspectratio
]{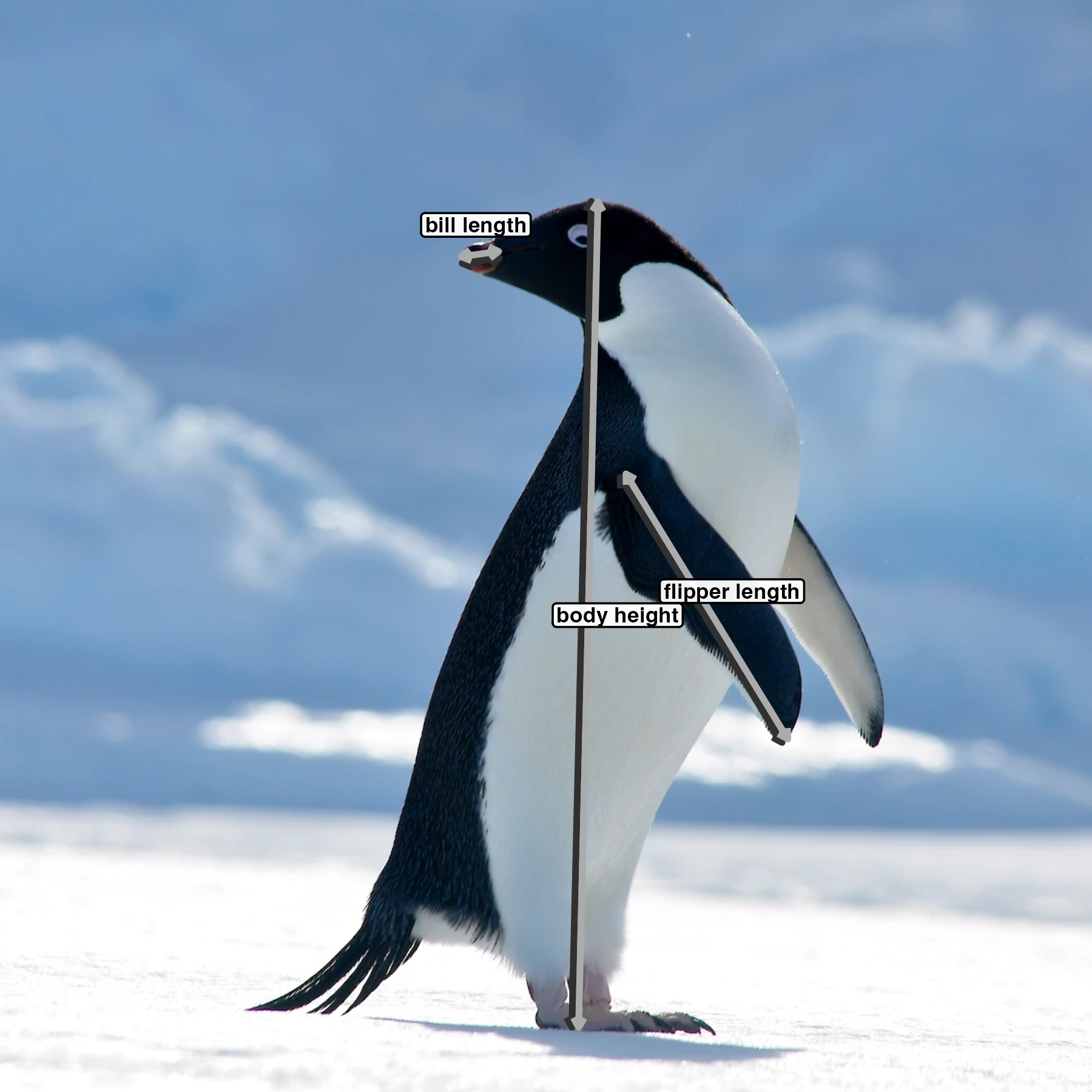}

\caption{Dual-stroke annotation of an Adélie penguin showing bill
length, flipper length, and body height. Photograph adapted from the
Australian Antarctic Division \citep{aad_penguin_photo}.}

\label{fig:penguin}

\end{figure}

A second example, shown in Figure~\ref{fig:penguin}, illustrates the
same design principle on an image containing both bright snow and dark
plumage. The paired representation remains visible across these contrasting regions without requiring different manually selected annotation colors for separate portions of the image.

% ============================================================
% FIGURE 5: Penguin
% ============================================================

\FloatBarrier

% ------------------------------------------------------------
\subsection{Paths and Mathematical Functions}
% ------------------------------------------------------------

The dual-stroke framework extends to continuous paths and mathematical
functions. The function \code{geom\_function\_dual()} provides a
contrast-aware counterpart to \code{geom\_function()}, allowing smooth
curves to remain distinguishable when they cross heterogeneous
background regions.

\begin{figure}[htbp]
\centering

\includegraphics[
   width=0.70\textwidth,
   keepaspectratio
 ]{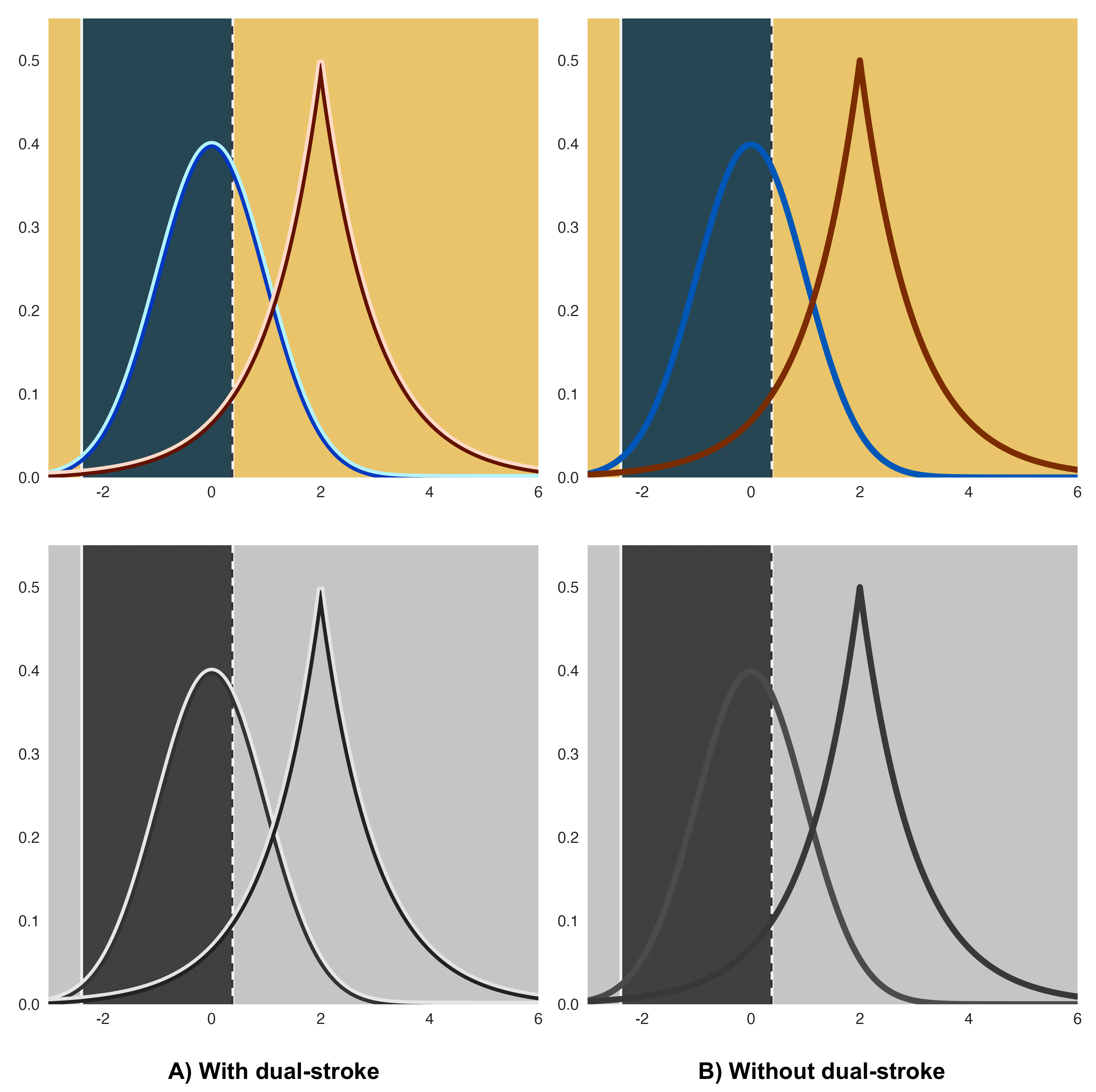}

\caption{Comparison of dual-stroke and single-stroke mathematical
functions across heterogeneous backgrounds and after grayscale
conversion.}

\label{fig:functions}

\end{figure}

Figure~\ref{fig:functions} compares conventional single-stroke and
dual-stroke mathematical functions over backgrounds containing both
light and dark areas. The same general rendering strategy is available
through \code{geom\_path\_dual()} for ordered trajectories whose
geometry is determined by data.

% ============================================================
% FIGURE 6: Mathematical functions
% ============================================================

\FloatBarrier

% ------------------------------------------------------------
\subsection{Regression Overlays}
% ------------------------------------------------------------

Regression overlays provide another setting in which local contrast can
vary substantially. A fitted line may pass through point clouds with
different colors, densities, or luminance, making a single foreground
color difficult to distinguish consistently.

\begin{figure}[htbp]
\centering

 \includegraphics[
   width=0.70\textwidth,
   keepaspectratio
 ]{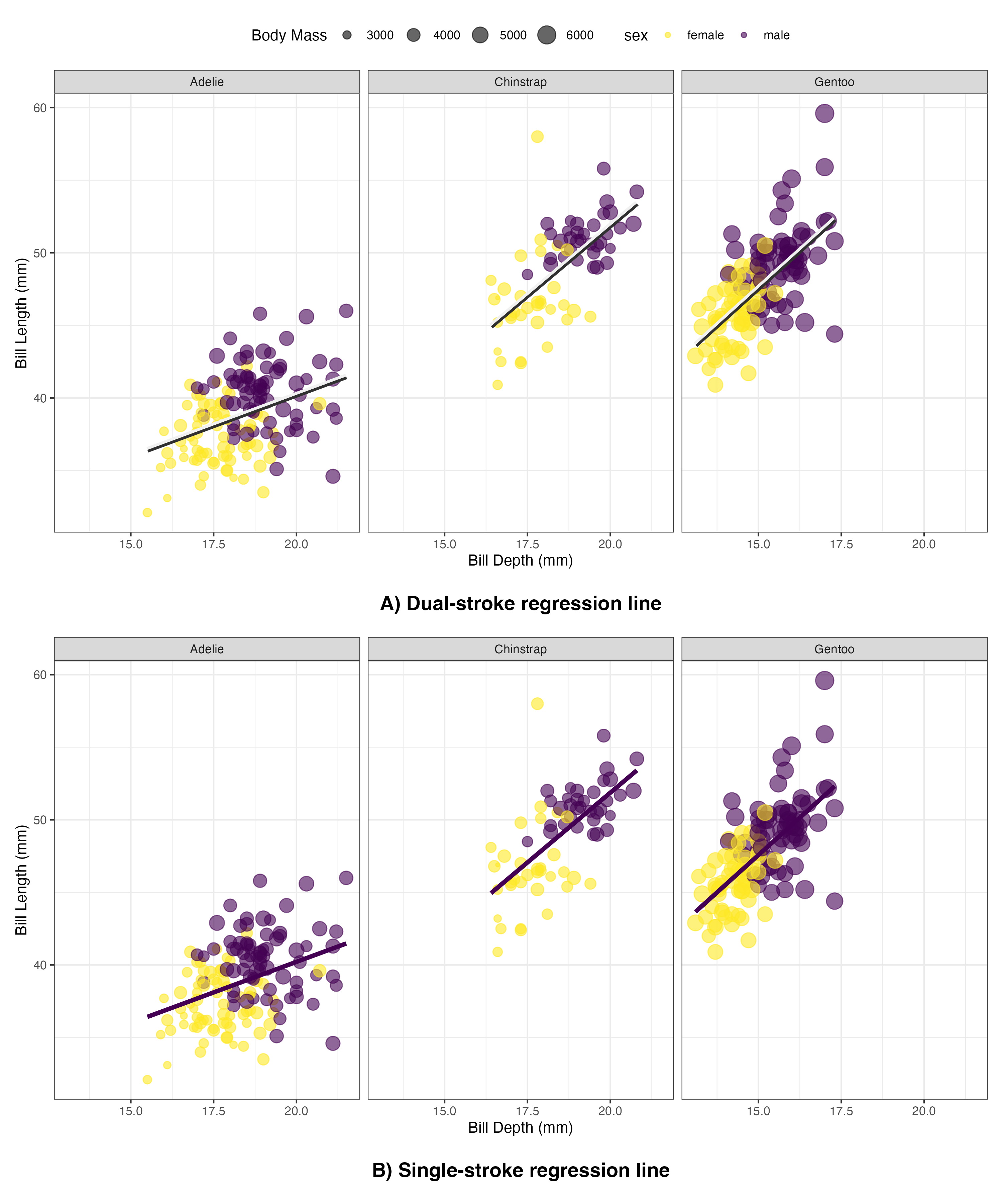}

\caption{Comparison of regression overlays using dual-stroke and
single-stroke rendering on the Palmer Penguins data.}

\label{fig:regression}

\end{figure}

The Palmer Penguins example in Figure~\ref{fig:regression} uses data
from the \pkg{palmerpenguins} package \citep{Horst2020} to compare a
standard regression overlay with the dual-stroke representation
produced by \code{geom\_lm\_dual()}. The paired light and dark
components improve the visibility of the fitted relationship where the
background and point colors vary.

% ============================================================
% FIGURE 7: Regression comparison
% ============================================================

\FloatBarrier

% ------------------------------------------------------------
\subsection{Contrast-Aware Text}
% ------------------------------------------------------------

Text labels require a different treatment because duplicating glyph
geometry can reduce legibility. The function
\code{geom\_text\_contrast()} instead evaluates the background
associated with a label and selects a light or dark foreground variant
using the selected contrast criterion.

\begin{figure}[htbp]
\centering

\includegraphics[
  width=0.70\textwidth,
  keepaspectratio
]{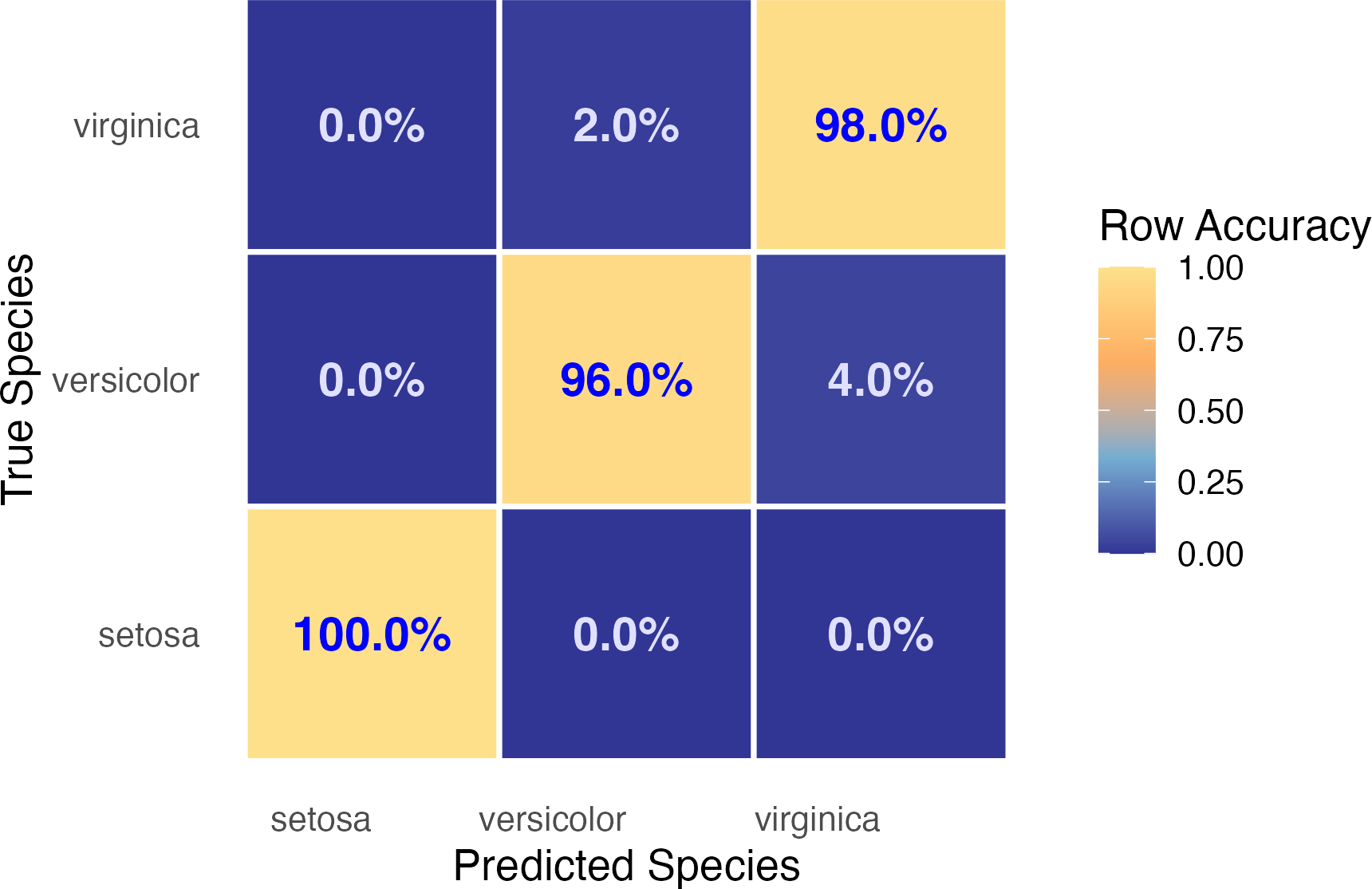}

\caption{Confusion matrix for a linear discriminant analysis classifier
applied to the iris data, with labels selected using
\code{geom\_text\_contrast()}.}

\label{fig:confusion}

\end{figure}

Figure~\ref{fig:confusion} illustrates this behavior using a confusion
matrix derived from the iris data originally analyzed by
\citet{Fisher1936}. The cell backgrounds span a range of intensities,
making a single fixed text color suboptimal across the matrix.

% ============================================================
% FIGURE 8: Confusion matrix
% ============================================================

\begin{figure}[htbp]
\centering

\includegraphics[
  width=0.92\textwidth,
  keepaspectratio
]{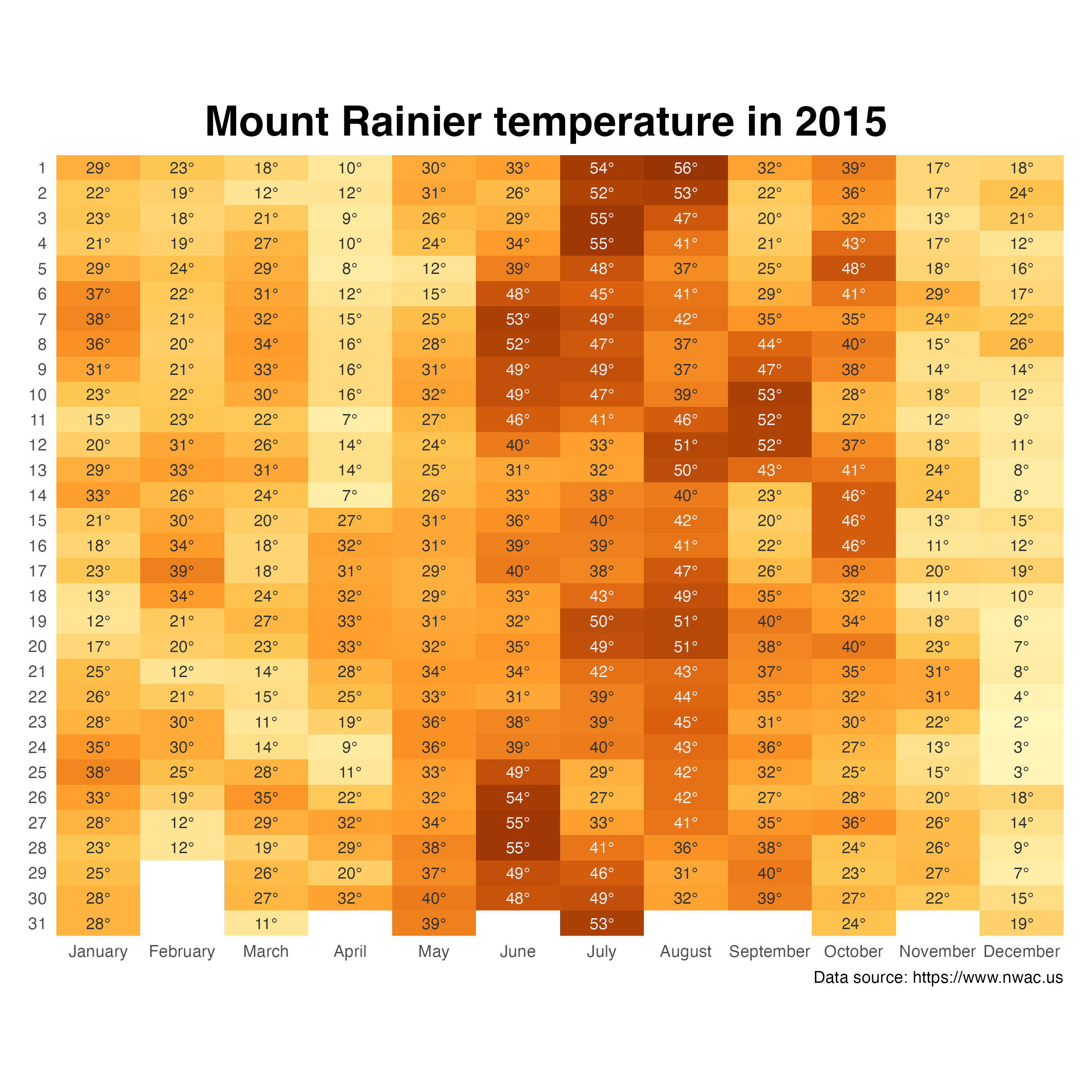}

\caption{Temperature heatmap with automatically selected
contrast-aware text labels.}

\label{fig:heatmap}

\end{figure}

A larger heatmap provides a more demanding example because many labels
are displayed across a continuous range of background colors. Using
temperature data from the Datylon heatmap example
\citep{DatylonHeatmap}, Figure~\ref{fig:heatmap} illustrates
automatically selected contrast-aware text colors across the display.

% ============================================================
% FIGURE 9: Heatmap
% ============================================================

\FloatBarrier

% ------------------------------------------------------------
\subsection{Highlight Palettes}
% ------------------------------------------------------------

The function \code{highlight\_colors()} addresses a complementary
visualization task: selecting emphasis colors that remain distinguishable
against a fixed background and base color while also remaining
perceptually separated from one another.

\begin{figure}[htbp]
\centering

\includegraphics[
  width=0.92\textwidth,
  keepaspectratio
]{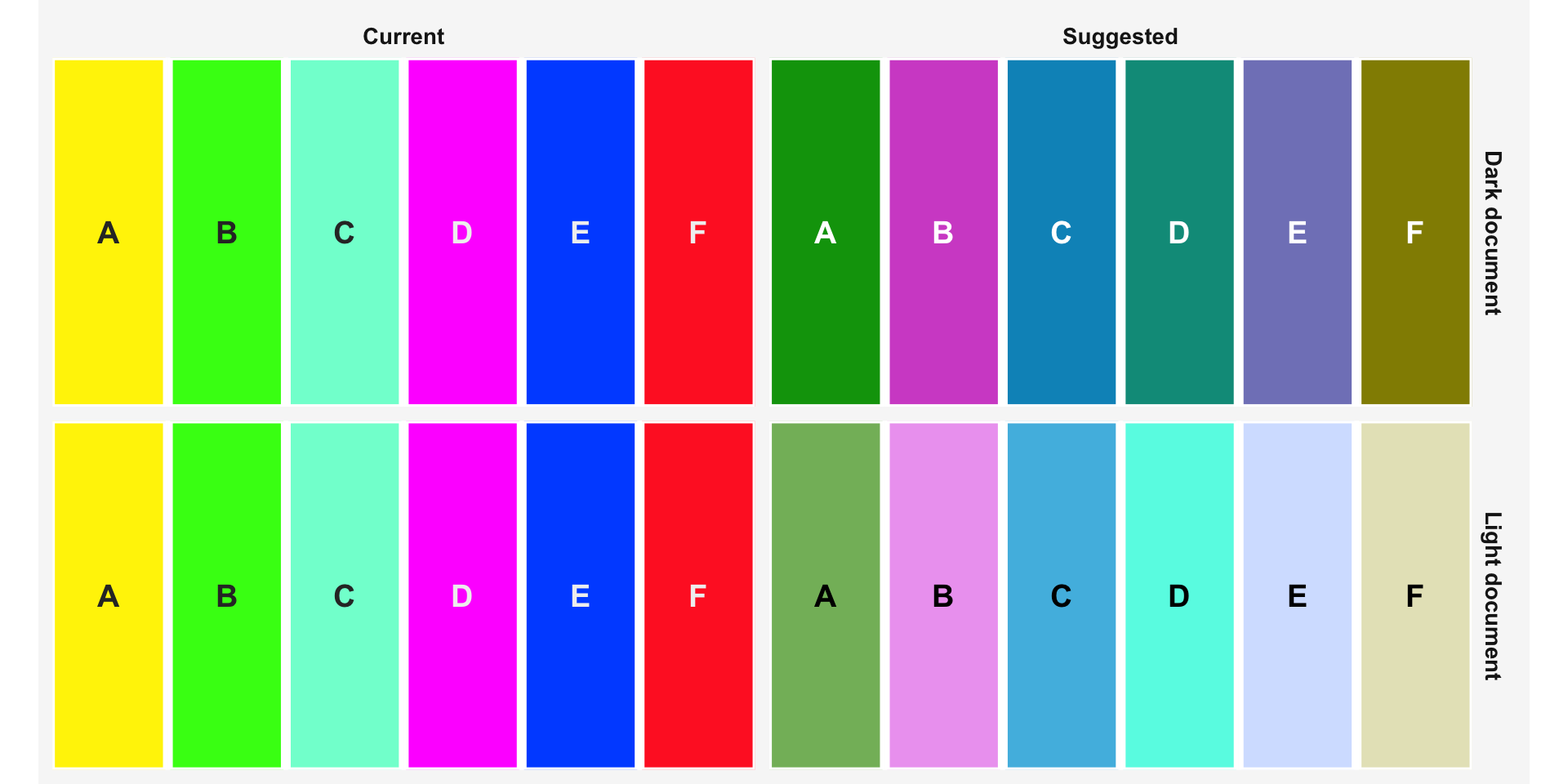}

\caption{Comparison of conventional document highlight colors and
context-aware palettes generated by \code{highlight\_colors()}.}

\label{fig:highlights}

\end{figure}

Unlike a fixed highlight palette, generated colors depend on the visual
context supplied to the function. Figure~\ref{fig:highlights} compares
the default text highlight colors available in Microsoft Word
\citep{MicrosoftWord2024} with context-aware palettes generated using
\code{highlight_colors()}.

% ============================================================
% FIGURE 10: Highlight palettes
% ============================================================

\FloatBarrier

% ------------------------------------------------------------
\subsection{Scientific Image Annotation}
% ------------------------------------------------------------

Scientific images provide a challenging setting for annotation because
local background characteristics can change sharply across short
distances. Images derived from microscopy, remote sensing, forensic
imaging, and other scientific instruments may contain both high- and
low-luminance regions together with texture that competes visually with
overlaid annotations.

\begin{figure}[htbp]
\centering

\includegraphics[
  width=0.70\textwidth,
  keepaspectratio
]{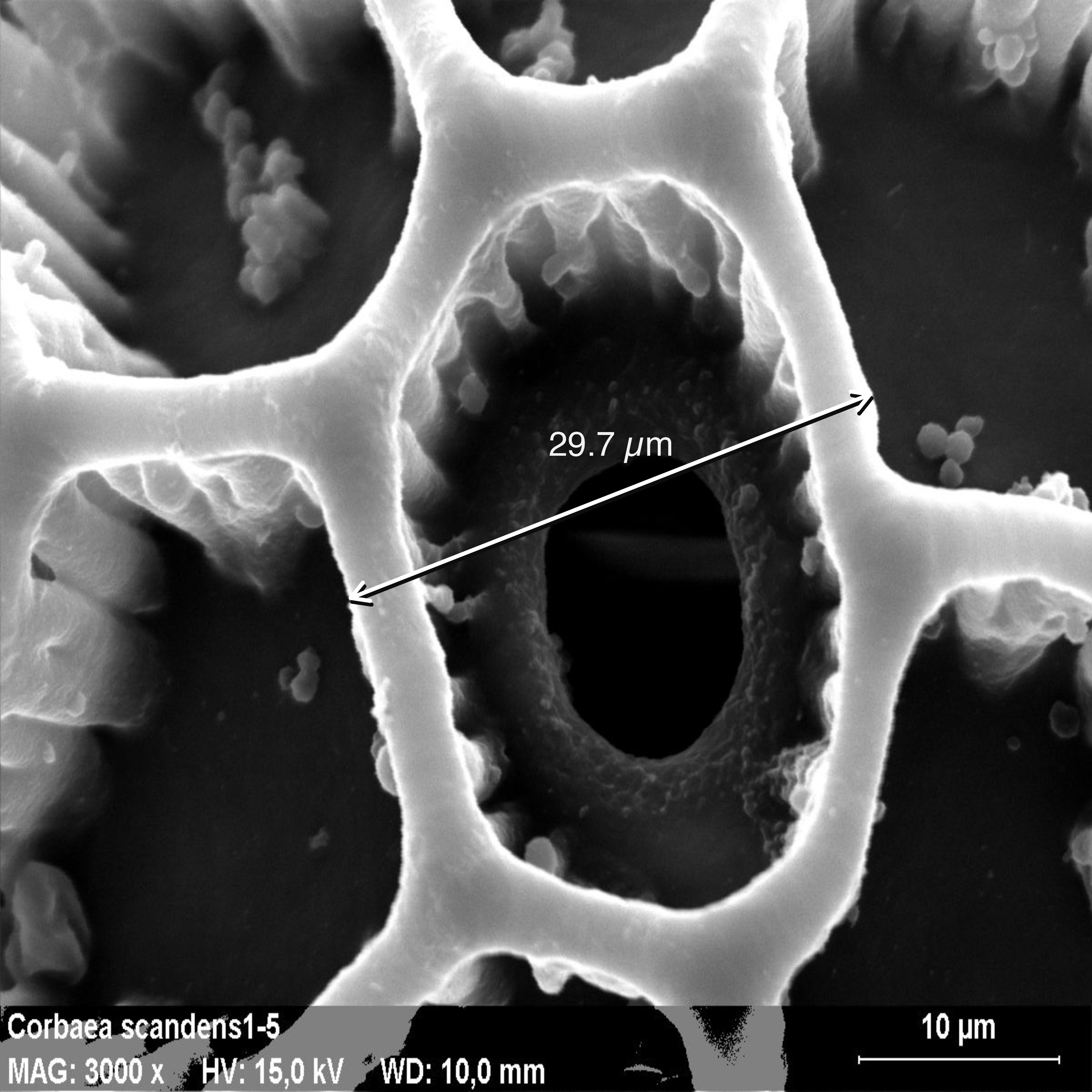}

\caption{Contrast-aware annotation applied to a scanning electron
micrograph of \emph{Cobaea scandens} pollen. The dual-stroke measurement
annotation remains distinguishable across regions of varying luminance
and texture. SEM micrograph adapted from Majaura
\citep{Majaura2007}, licensed under CC BY-SA 3.0.}

\label{fig:micro}

\end{figure}

Figure~\ref{fig:micro} provides an additional example using a scanning
electron micrograph of \emph{Cobaea scandens} pollen
\citep{Majaura2007}. The measurement annotation crosses regions with
substantial variation in luminance and texture, illustrating how the
dual-stroke representation can maintain visual separation from the
underlying image.

Together with the preceding scientific-image examples, this
demonstrates how the dual-stroke framework can be applied without
requiring separate manually selected annotation styles for individual
regions. The same general contrast-aware principles can therefore
support both statistical graphics and image-based scientific figures
within a reproducible R workflow.

% ============================================================
% FIGURE 11: Scientific image annotation
% ============================================================

\FloatBarrier

% ============================================================
\section{Simulation-Based Evaluation}
\label{sec:simulation}
% ============================================================

The application examples demonstrate the behavior of \pkg{ggtwotone} in
representative visualization settings. To complement these examples, we
conducted a simulation-based evaluation of the contrast properties of
three principal components of the package: dual-stroke rendering,
contrast-aware text, and highlight palette generation.

The objective of the simulation was not to model human visual
perception directly, but to evaluate whether the contrast-aware
procedures behave consistently across a broad range of controlled
background conditions.

% ------------------------------------------------------------
\subsection{Simulation Design}
% ------------------------------------------------------------

We generated 5,256 unique background colors consisting of a systematic
grayscale sequence together with randomly sampled RGB colors. For each
background, relative luminance and WCAG contrast ratios were calculated.
A base annotation color of \code{\#777777} was used throughout the
primary simulation.

For dual-stroke annotations, the contrast-adjustment engine generated a
light and dark color pair from the base color. The effective contrast
against background $B_j$ was evaluated as

\begin{equation}
M_j
=
\max
\left\{
\mathcal{C}(c_L,B_j),
\mathcal{C}(c_D,B_j)
\right\}.
\label{eq:simulation-dual}
\end{equation}

The dual-stroke representation was compared with three fixed
single-color strategies: black, white, and the unadjusted base color. A
contrast ratio of 3:1 was used as a reference criterion for graphical
annotations.

For contrast-aware text, the same backgrounds were used to compare
fixed black, fixed white, the unadjusted base color, and adaptive
foreground selection. For each background, the adaptive strategy
selected the light or dark candidate providing greater contrast. A
contrast ratio of 4.5:1 was used as a reference criterion for ordinary
text.

Finally, the behavior of \code{highlight\_colors()} was evaluated using
1,000 simulated background/base-color combinations for each requested
palette size

\[
n\in\{2,3,4,5\}.
\]

For each generated palette, we recorded whether the requested number of
colors was returned, the minimum contrast against the background and
base color, the minimum pairwise CIEDE2000 color difference, and the
minimum pairwise hue separation.

Because the package permits constraint relaxation when a complete
palette cannot satisfy all requested criteria simultaneously, these
results characterize the practical behavior of the implemented
palette-generation procedure rather than strict feasibility under the
initial constraints.

% ------------------------------------------------------------
\subsection{Dual-Stroke Contrast Stability}
% ------------------------------------------------------------

The dual-stroke representation substantially improved worst-case
contrast relative to each fixed single-color strategy.
Table~\ref{tab:dual-simulation} summarizes the distribution of WCAG
contrast ratios across the 5,256 simulated backgrounds.

\begin{table}[htbp]
\centering

\caption{Contrast performance of single-color and dual-stroke rendering
across simulated backgrounds.}

\label{tab:dual-simulation}

\begin{tabular}{lrrrr}

\toprule
Method & Minimum & Median & Mean & Pass rate (\%)\\
\midrule

Single base color & 1.00 & 1.76 & 1.97 & 14.1\\
Single black      & 1.00 & 5.99 & 7.12 & 80.5\\
Single white      & 1.00 & 3.51 & 4.68 & 58.1\\
\pkg{ggtwotone} dual & 3.42 & 5.87 & 6.56 & 100.0\\

\bottomrule

\end{tabular}

\end{table}

All three fixed-color strategies reached a minimum contrast ratio of
1.00 under at least one simulated background. In comparison, the
dual-stroke representation maintained a minimum effective contrast of
3.42. Its median contrast was 5.87, compared with 1.76 for the
unadjusted base color, 5.99 for black, and 3.51 for white.

Fixed black produced a slightly larger mean contrast ratio than the
dual-stroke representation, 7.12 versus 6.56. This reflects its very
high contrast against light backgrounds rather than consistently better
performance. Its contrast approached 1.00 against sufficiently dark
backgrounds.

The dual representation therefore does not necessarily maximize average
contrast. Instead, under the evaluated conditions, it reduces
low-contrast failures by providing complementary light and dark
components.

% ------------------------------------------------------------
\subsection{Adaptive Text Contrast}
% ------------------------------------------------------------

Adaptive foreground selection showed a similar reduction in
low-contrast cases for text annotations.
Table~\ref{tab:text-simulation} compares the adaptive strategy with
fixed black, fixed white, and the unadjusted base color.

\begin{table}[htbp]
\centering

\caption{Contrast performance of fixed and adaptive text colors across
simulated backgrounds.}

\label{tab:text-simulation}

\begin{tabular}{lrrrr}

\toprule
Method & Minimum & Median & Mean & Pass rate (\%)\\
\midrule

Fixed base color & 1.00 & 1.76 & 1.97 & 0.15\\
Fixed black      & 1.00 & 5.99 & 7.12 & 64.6\\
Fixed white      & 1.00 & 3.51 & 4.68 & 37.0\\
\pkg{ggtwotone} adaptive & 4.50 & 6.32 & 7.03 & 100.0\\

\bottomrule

\end{tabular}

\end{table}

Across the 5,256 simulated backgrounds, the adaptive strategy maintained
a minimum contrast ratio of 4.50 and satisfied the specified 4.5:1
criterion in all evaluated cases. In comparison, fixed black satisfied
the criterion for 64.6\% of backgrounds and fixed white for 37.0\%.
The unadjusted base color satisfied the criterion for approximately
0.15\% of the simulated backgrounds.

As in the dual-stroke evaluation, maximizing mean contrast was not the
primary objective. Fixed black produced a slightly higher mean contrast
than adaptive selection, 7.12 versus 7.03, but its minimum contrast fell
to 1.00. Adaptive selection instead maintained the specified criterion
across the range of backgrounds considered in the simulation.

% ------------------------------------------------------------
\subsection{Highlight Palette Behavior}
% ------------------------------------------------------------

The highlight-palette simulation examined how the perceptual properties
of generated palettes changed as the number of requested colors
increased. Table~\ref{tab:highlight-simulation} summarizes the results
from 1,000 simulated background/base-color combinations for each
palette size.

\begin{table}[htbp]
\centering

\caption{Behavior of highlight palette generation as the requested
number of colors increases.}

\label{tab:highlight-simulation}

\small

\begin{tabular}{rrrrr}

\toprule
Requested &
Complete &
Median min. &
Median min. &
Median min.\\

colors &
palettes (\%) &
background contrast &
$\Delta E_{00}$ &
hue separation\\

\midrule

2 & 63.8 & 4.58 & 35.4 & 145.6\\
3 & 63.8 & 4.53 & 28.1 & 79.1\\
4 & 63.5 & 4.53 & 22.7 & 43.2\\
5 & 63.5 & 4.52 & 19.5 & 27.4\\

\bottomrule

\end{tabular}

\end{table}

The proportion of simulations returning the complete requested palette
remained relatively stable across palette sizes, ranging from 63.5\% to
63.8\%. Median minimum background contrast also remained comparatively
stable, decreasing only slightly from 4.58 for two-color palettes to
4.52 for five-color palettes.

In contrast, pairwise perceptual separation decreased as additional
highlight colors were requested. The median minimum CIEDE2000 distance
declined from 35.4 for two-color palettes to 19.5 for five-color
palettes, while median minimum hue separation decreased from
approximately $145.6^\circ$ to $27.4^\circ$.

These results illustrate the increasing difficulty of maintaining large
pairwise perceptual separation as more colors are selected from the
available color space.

Because the simulation allowed the package's constraint-relaxation
mechanism, the reported values should not be interpreted as strict
satisfaction rates for the initially requested contrast and separation
thresholds. Instead, they characterize the trade-off produced by the
implemented fallback procedure when all requested constraints cannot be
satisfied simultaneously.

Taken together, the simulation results support the intended behavior of
the contrast-aware framework while also identifying practical
trade-offs. Dual-stroke rendering increased worst-case contrast relative
to fixed single-color annotations under the evaluated conditions, and
adaptive text selection avoided low-contrast cases in the evaluated
background set. Highlight palette generation maintained comparatively
stable background contrast as palette size increased, although pairwise
perceptual and hue separation decreased as more colors were requested.

% ============================================================
\section{Discussion}
% ============================================================

The results demonstrate an important distinction between maximizing
average contrast and maintaining robust annotation visibility across
heterogeneous backgrounds. A fixed foreground color can perform very
well over some backgrounds while becoming difficult to distinguish over
others. This behavior was particularly apparent for fixed black in the
simulation. Although black produced a slightly higher mean contrast than
the dual-stroke strategy, its minimum contrast fell to 1.00. The
dual-stroke approach instead improved the worst-case behavior by
combining complementary foreground components.

This distinction is central to the motivation for \pkg{ggtwotone}. The
package is not designed to identify a single color that maximizes
average contrast across an entire figure. Rather, it reduces dependence
on a single foreground--background relationship by incorporating
contrast-aware redundancy or adaptation into the annotation itself.
This is particularly useful for annotations that traverse images,
heatmaps, dense point clouds, or other graphical regions with changing
luminance.

The adaptive text results illustrate a related principle. Fixed black
and fixed white each performed well for subsets of the simulated
backgrounds, but neither was robust across the complete evaluated
background set. Selecting between light and dark foreground candidates
according to the local background avoided these low-contrast cases in
the simulation. Importantly, this result should be interpreted as an
evaluation of contrast metrics rather than as direct evidence of human
reading performance. The present study did not include a perceptual user
experiment.

The highlight-palette results demonstrate a different trade-off. While
background contrast remained comparatively stable as the requested
palette size increased, pairwise perceptual and hue separation declined.
This behavior reflects a general constraint of selecting multiple colors
within a finite perceptual color space while simultaneously imposing
contrast requirements. The relaxation mechanism in
\code{highlight\_colors()} provides a practical response to this
constraint, but it also means that requested separation thresholds
should be interpreted as target constraints rather than universal
guarantees.

The framework has several practical advantages. First, it integrates
directly with the \pkg{ggplot2} layer system, allowing contrast-aware
annotations to be introduced without requiring a separate visualization
grammar. Second, a shared contrast engine provides a consistent basis
for several annotation types. Third, the approach makes contrast-related
choices reproducible because the rules used to generate foreground
colors can be encoded directly in plotting code rather than adjusted
manually after figure construction.

Several limitations should also be recognized. WCAG and APCA provide
quantitative contrast criteria, but numerical contrast measures are not
complete models of human visual perception. Visibility can also depend
on line width, font size, spatial frequency, texture, display
conditions, visual acuity, and surrounding graphical structure. The
simulation therefore evaluates algorithmic contrast behavior rather
than perceptual superiority.

In addition, the present evaluation focuses primarily on controlled
color backgrounds. Real scientific graphics can contain complex
textures, transparency, anti-aliasing, overlapping graphical objects,
and spatially varying visual density. These properties may affect
perceived annotation visibility in ways not represented by a
foreground--background contrast ratio alone.

Dual-stroke rendering can also increase the visual weight of an
annotation relative to a conventional single stroke. While this
redundancy supports visibility, it may not be desirable for every
visualization. The approach is most useful when maintaining annotation
visibility across heterogeneous backgrounds is more important than
preserving the minimal appearance of a single-color line.

Future work could evaluate the framework through controlled perceptual
experiments in which participants identify or interpret annotations
under varying background, display, and viewing conditions. Such studies
would complement the present metric-based simulation and could quantify
effects on detection accuracy, reading time, and interpretation.

Additional work could examine performance under color-vision
deficiencies, projection conditions, grayscale printing, and different
display technologies. The framework could also be extended to
additional \pkg{ggplot2} annotation types and to alternative perceptual
models as accessibility and contrast standards continue to evolve.

% ============================================================
\section{Conclusion}
% ============================================================

This paper introduced \pkg{ggtwotone}, an R package for incorporating
contrast-aware annotation into \pkg{ggplot2} graphics. The package
addresses a recurring visualization problem: a single annotation color
that is clearly visible in one part of a figure may lose contrast as the
local background changes.

The proposed framework addresses this problem through three related
strategies. Dual-stroke geoms combine complementary light and dark
components for line-based annotations, \code{geom\_text\_contrast()}
adapts foreground text color to local background conditions, and
\code{highlight\_colors()} generates context-dependent highlight
palettes subject to contrast and perceptual-separation criteria. These
components are supported by a common contrast-adjustment framework and
integrate with the existing \pkg{ggplot2} layer system.

Simulation results showed that the dual-stroke representation improved
worst-case contrast relative to fixed single-color alternatives under
the evaluated conditions, while adaptive text selection maintained the
specified WCAG contrast criterion across the simulated backgrounds.
The highlight-palette evaluation further demonstrated the trade-off
between maintaining background contrast and preserving pairwise
perceptual separation as the number of requested colors increases.

More broadly, \pkg{ggtwotone} treats annotation visibility as a property
that can be incorporated into the visualization process rather than
addressed solely through manual post-processing. By making
contrast-related decisions explicit and reproducible, the framework
provides a practical foundation for developing statistical graphics
that remain interpretable across heterogeneous visual contexts.

\clearpage

% ============================================================
\section*{Software and Reproducibility}
% ============================================================

The \pkg{ggtwotone} package is distributed through the Comprehensive
R Archive Network (CRAN). Source code, documentation, examples, and
development materials are also available through the project's GitHub
repository.

\begin{itemize}
  \item CRAN package:
  \url{https://cran.r-project.org/package=ggtwotone}

  \item Source repository:
  \url{https://github.com/bwanniarachchige2/ggtwotone}

  \item Package documentation:
  \url{https://bwanniarachchige2.github.io/ggtwotone/}
\end{itemize}

% ============================================================
\section*{Author Information}
% ============================================================

\noindent
\textbf{Beenu Sareena}\\
Department of Statistics\\
University of Nebraska--Lincoln\\
Nebraska, USA\\
ORCID: 0009-0002-7611-5499\\
Email: \texttt{79365709@nebraska.edu}

\vspace{0.5em}

\noindent
\textbf{Heike Hofmann}\\
Department of Statistics\\
University of Nebraska--Lincoln\\
Nebraska, USA\\
ORCID: 0000-0001-6216-5183

\clearpage

% ============================================================
% References
% ============================================================

\bibliographystyle{plainnat}
\bibliography{references}

\end{document}